\documentclass[aps,prl,twocolumn,showpacs,superscriptaddress]{revtex4-1}

\usepackage{verbatim}
\usepackage{epsfig}
\usepackage{amssymb}
\usepackage{amsmath}
\usepackage{graphicx}
\def\Planck{\textit{Planck}}
\def\planck{\textit{Planck}}

\newcommand{\vn}{\hat{\mbox{n}}}
\newcommand{\vv}{\mathbf{v}}
\newcommand{\vrv}{\mathbf{r}}

\newcommand{\sevem}{{\tt SEVEM}}

\newcommand{\galaxy}{{\tt GALAXY}}
\newcommand{\pksz}{p_{\rm kSZ}(r)}
\newcommand{\wTv}{w^{\delta T\,v^{\rm rec}}(r)}

\usepackage{multirow}

\def\araa{{ARA\&A}}             
\def\apj{{ApJ}}                 
\def\apjl{{ApJ}}                
\def\apjs{{ApJS}}               
\def\apss{{Ap\&SS}}             
\def\aap{{A\&A}}                

\def\mnras{MNRAS}             
\def\nat{{Nature}}              

\providecommand{\sorthelp}[1]{}

\begin{document}
\title{Evidence of the missing baryons from the kinematic Sunyaev-Zeldovich effect in \Planck\ data}
\date{Received 18 April 2015; published November 3rd, 2015}

\author{Carlos~Hern\'andez--Monteagudo}
\affiliation{Centro de Estudios de F\'\i sica del Cosmos de Arag\'on (CEFCA), Plaza San Juan, 1, planta 2, E-44001,Teruel, Spain}
\author{Yin-Zhe~Ma}
\affiliation{
Astrophysics and Cosmology Research Unit, School of Chemistry and Physics, University of KwaZulu-Natal, Durban, South Africa}
\affiliation{Jodrell Bank Centre for Astrophysics, School of Physics and Astronomy, The University of Manchester, Manchester M13 9PL, UK}
\author{Francisco~S.~Kitaura}
\affiliation{Leibniz-Institut f\"ur Astrophysik Potsdam (AIP), An der Sternwarte 16, D-14482 Potsdam, Germany}
\author{Wenting~Wang}
\affiliation{Institute for Computational Cosmology, University of Durham, South Road, Durham, DH1 3LE, UK}
\author{Ricardo~G\'enova--Santos}
\affiliation{Instituto de Astrof\'{\i}sica de Canarias, C/V\'{\i}a L\'{a}ctea s/n, E-38205 La Laguna, Tenerife, Spain}
\affiliation{Departamento de Astrof\'{\i}sica, Universidad de La Laguna (ULL), 38206 La Laguna, Tenerife, Spain}
\author{Juan~Mac\'\i as-P\'erez}
\affiliation{Laboratoire de Physique Subatomique et Cosmologie, Universit\'{e} Grenoble-Alpes, CNRS/IN2P3, 53, rue des Martyrs, 38026 Grenoble Cedex, France}
\author{Diego~Herranz}
\affiliation{Instituto de F\'{\i}sica de Cantabria (CSIC-Universidad de Cantabria), Avda. de los Castros s/n, E-39005, Santander, Spain}

\begin{abstract}
We estimate the amount of the {\it missing baryons} detected by the \Planck\ measurements of the cosmic microwave background in the direction of Central Galaxies (CGs) identified in the Sloan galaxy survey. The peculiar motion of the gas inside and around the CGs unveils values of the Thomson optical depth $\tau_{\rm T}$ in the range $0.2$--$2\times 10^{-4}$,  indicating that the regions probed around CGs contain roughly half of the total amount of baryons in the Universe at the epoch where the CGs are found. If baryons follow dark matter, the measured $\tau_{\rm T}$s are compatible with the detection all the baryons existing inside and around the CGs.
\end{abstract}


\maketitle

{\it Introduction.} 
The interplay between baryons and dark matter is a key problem in cosmology and galaxy formation. Understanding the distribution of baryonic and dark matter in galaxies, groups, and clusters of galaxies is an essential step towards the full picture of how these objects form and evolve. It is well known \cite{Fukugita04,cenostriker2006,Bregman07} that only about $10\%$ of all baryons in the Universe reside in the form of stellar mass, while other $90\%$ resides in a diffuse, mostly undetected component, if it were not for recent UV spectroscopic measurements of un-collapsed, diffuse gas in the direction of certain QSOs, \cite[see, e.g.,][and references therein.]{shulletal12,tejosetal14} Nowadays there is an ongoing debate \citep[][]{Gupta12,Gatto13,planck2012-XI,Werk14,lebrunetal15,MillerBregman15} on whether a significant fraction of the latter component is present in the circumgalactic medium around halos or if instead most of the gas has been expelled or never accreted due to feedback processes like galactic winds or Active Galactic Nuclei activity. This connects the {\it missing baryon} issue with the complex problems of feedback and galaxy formation. Certainly a more complete census of the baryon distribution in the universe would be of great relevance in this context.

The kinematic Sunyaev-Zeldovich effect \citep[hereafter kSZ,][]{kSZ} describes the Doppler shift in frequency induced on photons of the cosmic microwave background (hereafter CMB) photons after they scatter off free electrons. This Thomson scattering induces frequency independent brightness temperature anisotropies in the CMB, which are given by \citep[][]{kSZveryfirst}
\begin{equation}
\delta T_{\rm kSZ}(\vn) = -T_{\rm 0}\,\int {\rm d}l\,\sigma_{\rm T} n_{\rm e} \left(\frac{\vv}{c} \cdot \vn\right)
  \simeq -T_{\rm 0}\,\tau_{\rm T} \left(\frac{\vv}{c} \cdot \vn\right).
\label{eq:tauapp1}
\end{equation}
In this expression, the integral $\tau_{\rm T} = \sigma_{\rm T} \,\int {\rm d}l\, n_{\rm e} $ is conducted along the line of sight (LOS) given by $\vn$. We have made the approximation that  the typical correlation length of LOS velocities (given by $\vv\cdot \vn$) is much larger than the density correlation length, in such a way that the LOS velocity term may be pulled out of the kSZ integral. This is justified by the results of \citet{Planck2015-CV}, who find a typical correlation length of peculiar velocities of $20$--$40\,h^{-1}$\,Mpc, well above the typical galaxy correlation length ($\sim 5\,h^{-1}$\,Mpc). The expression above also shows that the kSZ constitutes an integral over the electron momentum, independently of the temperature, and thus counts {\em all} electrons in the bulk flows, regardless they belong to collapsed structures or not. 

Large scale, bulk matter flows were detected via the kSZ effect firstly by \cite{Handetal2012} and more recently by \cite{Planck2015-CV}. We build upon the latter work and extract physical constraints on the amount of baryons contributing to the kSZ signal and the implications of those measurements in the problem of the missing baryons. As in \cite{Planck2015-CV}, we use the latest \Planck\ data release (DR2) available at the \Planck's Legacy Archive server \footnote{\Planck's Legacy Archive:  \\{\tt http://pla.esac.esa.int/pla/aio/planckProducts.html}}, the Central Galaxy Catalogue (hereafter CGC) obtained from the Sloan Digital Sky Survey SDSS/DR7 \cite{Abazajianetal2009}, and a mock catalogue of central galaxies obtained from the Millennium numerical simulation \cite{millennium05}, to which we henceforth  refer as the \galaxy\ mock catalogue. In this work we quote results for the \Planck\ \sevem\ CMB map, although similar results are obtained for all the other cleaning algorithms.  Our estimates of the kSZ temperature of Eq.~(\ref{eq:tauapp1}) are taken from \cite{Planck2015-CV} and are obtained after applying aperture photometry (hereafter AP) in the direction of the CGC members. We defer the reader to that paper for a more detailed description of the data under use, and a deeper discussion on the measurements of the kSZ, while in this work we concentrate upon the physical interpretation of such measurements.  In particular, we set constraints on $\tau_T$ and the missing baryons from our AP measurements of $\delta T_{\rm kSZ}$ in Eq.~\ref{eq:tauapp1}.

\begin{figure}
\centering
\includegraphics[width=9.cm]{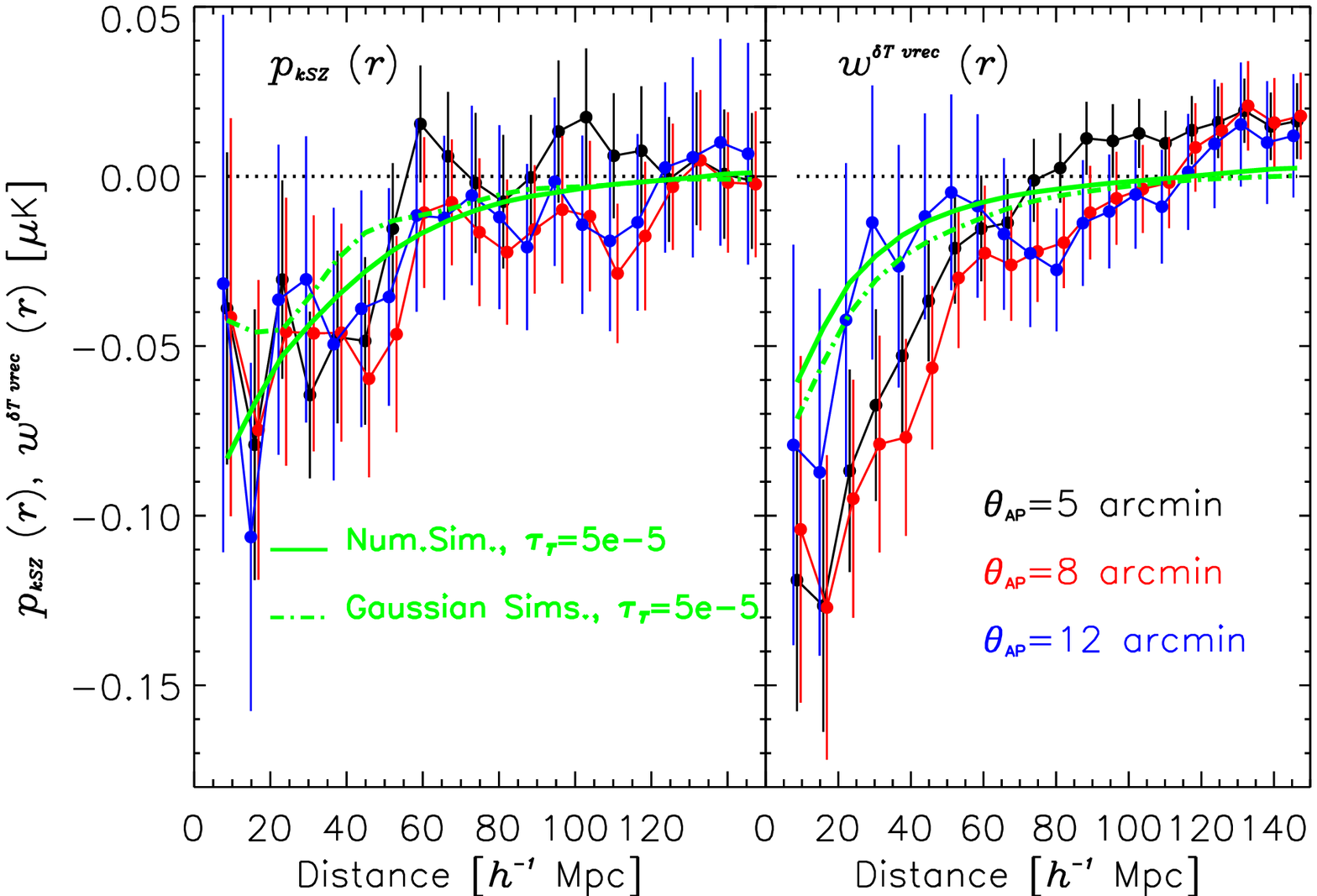}
\caption[fig:CFs]{ {\it Left panel: }Measured pairwise peculiar momenta under different aperture sizes for the \sevem\ map. {\it Right panel:} Measured $\wTv$ correlation function under different aperture radii ($\theta_{\rm AP}$) for the \sevem\ clean map. In both panels the green solid lines provide the expectation from the \galaxy\ mock catalogue, while the dash-dotted lines correspond to expectations found after averaging over 100 Gaussian simulations. In both cases, we assume $\tau_{\rm T}=5\times 10^{-5}$. 
}
\label{fig:CFs}
\end{figure}

\begin{figure}
\centering
\includegraphics[width=9.cm]{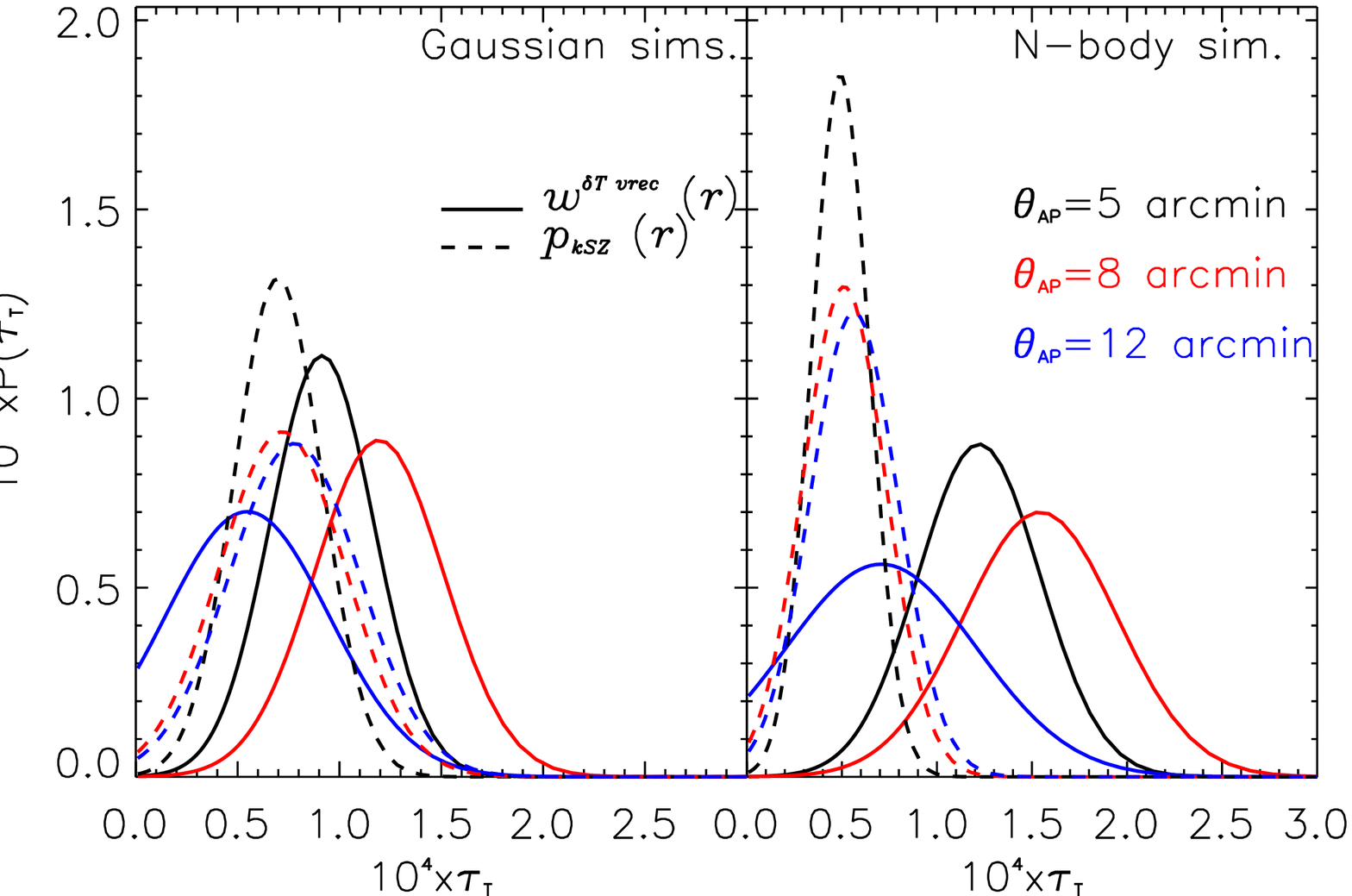}
\caption[fig:ptaus]{{\it Left panel:} Probability distribution for the optical depth $\tau_{\rm T}$ inferred from the $\wTv$ correlation function (solid lines) and the kSZ peculiar momenta (dashed lines), under different apertures, for the \sevem\ map. These estimates of $\tau_{\rm T}$ are obtained after fitting the data to the prediction inferred from the average of the Gaussian simulations. {\it Right panel: } Same as left panel, but using the \galaxy\ N-body prediction as a reference, rather than the Gaussian simulations. 
}
\label{fig:ptauss}
\end{figure}

{\it Methodology and results.} 
The two statistics presented in \cite{Planck2015-CV} yielding kSZ evidence are the kSZ pairwise peculiar momentum (hereafter $\pksz$) and the cross correlation function of the kSZ temperature and the recovered radial peculiar velocity (hereafter $\wTv$). The $\pksz$ reflects the gravitational infall in pairs of galaxies and is computed from the sum
\begin{equation}
\hat{p}_{\rm kSZ} (r) = -\frac{\sum_{i<j}(\delta T_{ i} - \delta T_{ j} )\,c_{ i,j}}{\sum_{ i<j} c_{ i,j}^2},
\label{eq:pksz1}
\end{equation}
where the weights $c_{ i,j}$ are given by \cite{ferreiraetal99}
\begin{equation}
c_{ i,j} = \hat{\vrv}_{ i,j} \cdot \frac{\hat{\vrv}_{ i}+\hat{\vrv}_{ j}}{2} = 
 \frac{(r_{ i}-r_{ j})(1+\cos\theta)}{2\sqrt{r_{ i}^2 + r_{ j}^2 - 2r_{ i}r_{ j}\cos\theta}}.
\label{eq:cweight}
\end{equation}
Following the convention of \cite{Handetal2012}, in this equation ${\vrv}_{ i}$ and ${\vrv}_{ j}$ are the vectors pointing to the positions of the $i$-th and $j$-th galaxies on the celestial sphere, $r_{ i}$ and $r_{ j}$ are the comoving distances to those objects, and ${\vrv}_{ i,j} = \vrv_{ i}-\vrv_{ j}$ refers to the distance vector for this pair of galaxies. The {\it hat} symbol ($\hat{\vrv}$) denotes a unit vector in the direction of $\vrv$, and $\theta$ is the angle separating $\hat{\vrv}_{ i}$ and $\hat{\vrv}_{ j}$. Note as well that $\hat{\vrv}_{i,j}$ refers to the direction of the difference vector, i.e. $(\vrv_i-\vrv_j)/|\vrv_i-\vrv_j|$. 
On the other hand, the $\wTv$ function is computed as the spatial correlation function between the measured kSZ temperature estimates and the linear theory predictions of the radial peculiar velocity of the CGs ($v^{\rm rec}$), $\wTv=\sum_{i,j} (\delta T_i\, v^{\rm rec}_j\,w_i\,w_j) / \sum_{i,j} w_i\,w_j$, where as before the $i,j$ sub-indices refer to CGs and $w_i,w_j$ to their weights. For both statistics, the $\delta T_j$s correspond to the AP estimates of the kSZ fluctuations of Eq.~(\ref{eq:tauapp1}). We refer again to \cite{Planck2015-CV} for details on how linear theory predictions of CG radial peculiar velocities are obtained. Figure~\ref{fig:CFs} displays the measured $\pksz$ and $\wTv$ functions under 5, 8, and 12\,arcmin aperture radii for the \sevem\ clean map of \Planck\ (very similar estimates are found for other CMB clean maps). In order to estimate the amount of free electrons giving rise to these signals, we compare these measurements with predictions obtained from {\it (i)} the \galaxy\ mock catalogue of central galaxies, and {\it (ii)} a suite of 100 Gaussian simulations of the matter density contrast field that is then inverted into a peculiar velocity field by means of the linearized continuity equation. The Gaussian simulations of the density contrast are generated from a dark matter linear power spectrum compatible with \planck's cosmology, and have no power at wavenumbers $k>0.15\,h$\,Mpc$^{-1}$ since those are regarded as non-linear scales. The first approach provides velocities for the halos hosting the CGs, while the second computes an estimate of the smooth, linear peculiar velocity in a region surrounding each CG. Following the approximation in Eq.~(\ref{eq:tauapp1}), green curves in Fig.~\ref{fig:CFs} provide predictions for $\pksz$ and $\wTv$ from the \galaxy\ mock (solid lines) and the Gaussian simulations (dash-dotted lines), with a choice of $\tau_{\rm T}=5\times 10^{-5}$ for display purposes. We find that, while for $\wTv$ the two predictions just differ in a $\sim 15$\,\% amplitude factor, for $\pksz$ they differ both in shape and amplitude. Since the kSZ is built upon all electrons present in the volume sampled by the aperture photometry, and not necessarily bound to the CGs, we expect the real signal to lie between the two predictions.

Following the template fit approach used in \cite{Planck2015-CV} that conducts minimum $\chi^2$ fits of the data to the predictions while assuming Gaussian and correlated errors, we obtain the density probabilities for $\tau_{\rm T}$ as given in Fig.~\ref{fig:ptauss}. When fitting to the predictions inferred from the \galaxy\ catalogue, we obtain $\tau_{\rm T}$ estimates from $\pksz$ measurements falling in the $(0.1$--$1.1)\times 10^{-4}$ range, in slight tension (between 1 and $2\,\sigma$ low) compared to estimates from $\wTv$. A better consistency among $\tau_{\rm T}$ estimates is found when fitting to the predictions provided by the Gaussian simulations (see left panel of Fig.~\ref{fig:ptauss}). The $\pksz$ and $\wTv$ measurements should yield similar $\tau_{\rm T}$ estimates,
and hence Fig.~\ref{fig:ptauss} suggests that the motion of electrons in and around CGs is better described by the Gaussian simulations, which we adopt in subsequent analyses. Note that this choice is not crucial for our discussion below.

\begin{figure}
\centering
\includegraphics[width=8.cm]{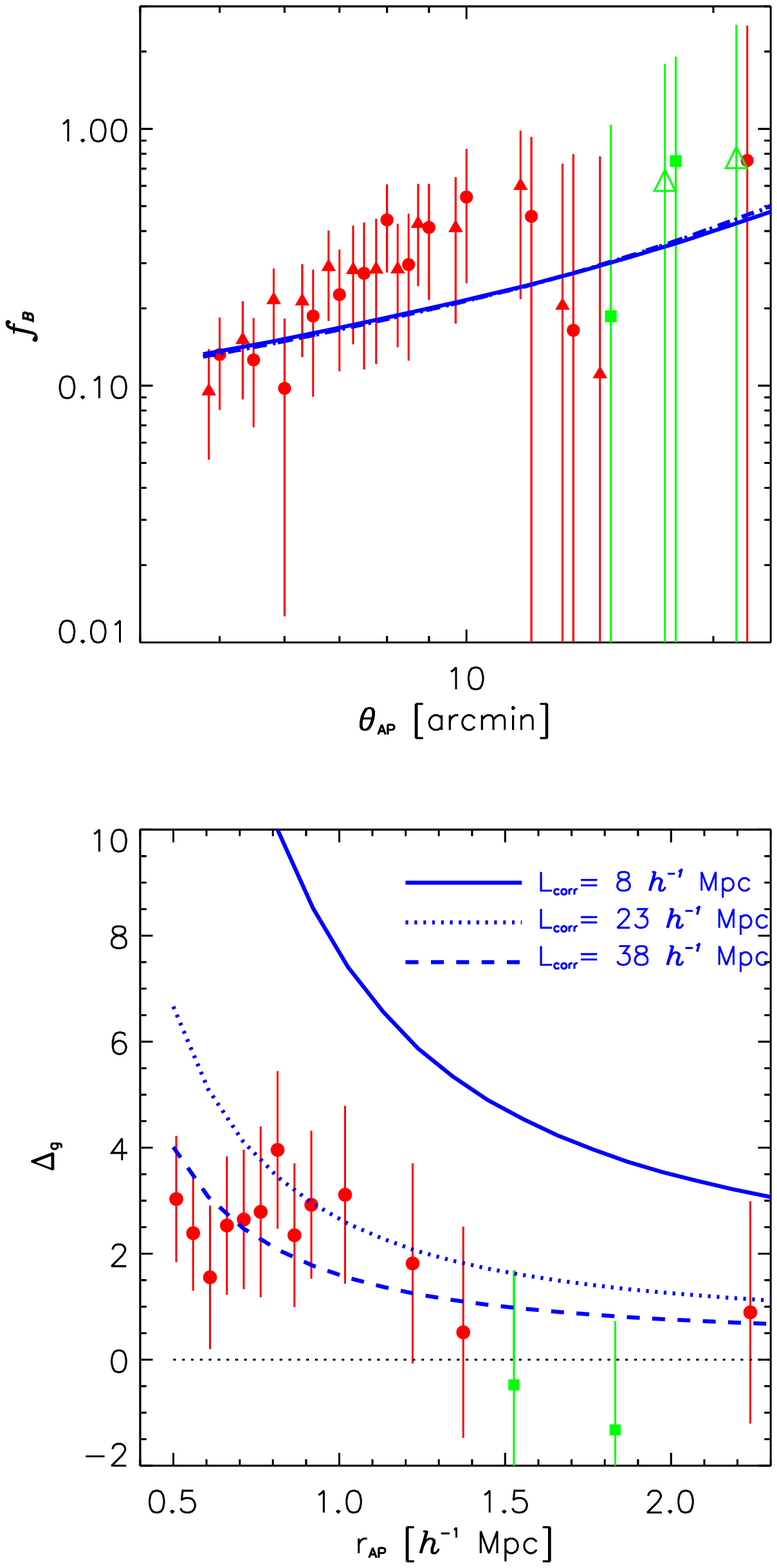}
\caption[fig:fb_Dg]{{\it Top panel:} Baryon fraction versus aperture angle, as inferred from $\wTv$ (filled circles and squares) and $\pksz$ (filled and empty triangles) measurements. Red (green) color denotes positive (negative) values. {\it Bottom panel:} Recovered profile of gas overdensity from $\wTv$ measurements and for $L_{\rm corr}=38\,h^{-1}$\,Mpc (filled symbols). In both panels, we compare our data to the predictions for dark matter \citep[as provided by][]{hayashiwhite08}, which are given by the blue lines. 
}
\label{fig:fb_Dg}
\end{figure}

 We next estimate the baryon fraction giving rise to the observed $\tau_{\rm T}$ amplitudes under different apertures.  For any line of sight throughout any CG, all electrons inside a cylinder whose base radius is given by the angular aperture ($\theta_{\rm AP}$) should contribute to the signal. The LOS depth of this cylinder corresponds to the comoving length of typical velocity correlation ($L_{\rm corr}$), and is suggested by Fig.~\ref{fig:CFs} to lie in the 20--40$\,h^{-1}$\,Mpc range (comoving units). We find however that the baryon fraction should not depend upon $L_{\rm corr}$ for any given aperture:
\[
\langle f_{\rm b} \rangle_z (\theta_{\rm AP}) =\left\langle \frac{\tau_{\rm T} (\theta_{\rm AP},z) / (\sigma_T {\bar n}_{\rm e}(z) L_{\rm corr}/(1+z)) }{1/f_{\rm vol}(z)}\right\rangle_{z} \approx
\]
\begin{equation}
\phantom{xxxxxx}
  \pi \frac{ \langle \tau_{\rm T}(\theta_{\rm AP},z)\rangle_z}{\sigma_{\rm T} {\bar n}_{\rm e,0}} \,\theta^2_{\rm AP}\,  \langle n_{\rm CG}(z) d_{\rm A}^2(z)\rangle_{z},
\label{eq:fb}
\end{equation}
where $f_{\rm vol} (z)\simeq n_{\rm CG}(z)\,\pi (r(z) \,\theta_{\rm AP})^2 L_{\rm corr}$ is the fraction of the comoving volume sampled by cylinders centred upon CGs, $L_{\rm corr}$ is the LOS depth of the cylinder in comoving units, ${\bar n}_{\rm e} (z)$ is the average physical electron number density at redshift $z$ (whereas the ``0" subscript denotes at present), $n_{\rm CG}(z)$ is the CG comoving number density, and $r(z)$ and $d_{\rm A}(z)$ are the comoving and angular distances to redshift $z$, respectively. The brackets $\langle ...\rangle_z$ denote redshift averages, and thus our previous measurements of $\tau_{\rm T}$ correspond to $\tau_{\rm T} (\theta_{\rm AP})= \langle \tau_{\rm T}(\theta_{\rm AP},z) \rangle_z$. Note that Eq.~(\ref{eq:fb}) holds as long as there is no overlap between cylinders along two different LOSs: we have verified that such overlap is negligible for apertures smaller than 13\,arcmin,  (it affects less than 1\,\% of pairs falling in the first distance bin at $\sim 9\,h^{-1}$\,Mpc, and this ratio is still smaller for the other distance bins). We remark that this definition of baryon fraction computes the ratio of detected baryons around  CGs to the {\em total} amount of baryons. Eq.~\ref{eq:fb} shows that this baryon fraction depends on the CGC completeness via the product of the average CG number density times the average $\tau_T$ measured in and around those galaxies.

The top panel of Fig.~\ref{fig:fb_Dg} shows that the baryon fraction increases with the aperture, amounting to 45--55\,\% of the total of all baryons for $\theta_{\rm AP} \sim 8$--$10\,$arcmin. For larger apertures, our measurements become compatible with noise and oscillate around zero. The blue lines provide the prediction for a scenario where the baryons trace perfectly the distribution of dark matter. These predictions follow the approach of \cite{hayashiwhite08}, which compute the dark matter -- central halo correlation function $\xi_{\rm h,m}(r) (\propto \delta \rho_{\rm m}(r))$ which is dependent of both mass and redshift of the halos, and requires some linear bias relation $b(M,z)$ that we adopt from  \cite{STbias99}. These predictions compute, as in Eq.~(\ref{eq:fb}), the ratio between the mean dark matter overdensity inside the cosmological volume sampled by the cylinders centred upon the CGs and the inverse of the fraction of the total cosmological volume sampled by the same cylinders. We thus need to integrate $\xi_{\rm h,m}(r)$ inside the cylinders, and as shown in the plot these predictions are independent of the choice of $L_{\rm corr}$. We can also incorporate the impact of our AP filter in our prediction,  after taking into account the redshift and mass distribution of our CG sample for each aperture $\theta_{\rm AP}$. We find that our measurements of the baryon fraction are compatible with this prediction, although slightly falling on the high amplitude side: results from $\wTv$ are given by filled red circles (filled green squares) when positive (negative), while filled red triangles (empty green triangles) correspond to positive (negative) $f_{\rm b}$ estimates inferred from the $\pksz$. A template fit to the dark matter prediction (given by the blue lines and performed as in Eqs.~11--12 in \cite{Planck2015-CV},  thus accounting for covariance among different apertures) yields an amplitude of $A_{\rm DM}=1.15\pm 0.30$ for baryon fraction estimates obtained from the $\wTv$, and $A_{\rm DM}=1.20\pm 0.22$ from $\pksz$ measurements of the baryon fraction. In both cases we compare these observables to the prediction from Gaussian simulations. As mentioned above, tension at the $1.6\,\sigma$ level arises when comparing fits of the two $\wTv,\,\pksz$ observables to the N-body \galaxy\ prediction: $A_{\rm DM}=1.49\pm0.37$ vs $A_{\rm DM}=0.85\pm0.15$ from $\wTv$ and $\pksz$ data, respectively. We remark that calibration errors in CMB maps would  impact these amplitudes while leaving the significance of the kSZ signal unchanged.   

Alternatively, in the bottom panel of Fig.~\ref{fig:fb_Dg} we compare the gas overdensity values inferred from the $\tau_{\rm T}$ measurements from our $\wTv$ observations and $L_{\rm corr}=38\,h^{-1}$\,Mpc (filled symbols) with the predictions of dark matter according to \cite{hayashiwhite08} (blue dot-dashed line), for different choices $L_{\rm corr}$. In this case, the conversion between angular apertures and transversal distances accounts for the redshift distribution of the CG sources. Values for gas overdensity are obtained from $\tau_{\rm T} (\theta_{\rm AP})$ estimates via
\begin{equation}
\Delta_{\rm g} (r_{\rm AP})  \approx 
\frac{\langle \tau_{\rm T} (\theta_{\rm AP}, z) \rangle_z}{\langle \sigma_{\rm T} {\bar n}_{\rm e}(z) L_{\rm corr}/(1+z)\rangle_z}.
\label{eq:Dg}
\end{equation}
We find again good agreement with the dark matter prediction, finding no hints for feedback that is supposed to deplete gas from the inner halo regions. Similar results are obtained from $\pksz$ observations.

{\it Discussion.} Observations of metallic lines in the X-ray in the direction of high energy sources provide information about the gas density in the so called circumgalactic medium of the Milky Way \cite{Gupta12,Gatto13,Werk14,MillerBregman15}. While some authors claim to have found evidence for all the baryons expected in our halo \cite{Gupta12}, some other authors seem to find only between $10$ to $50$\,\% of the expected amount of baryons \cite{Gatto13,Werk14,MillerBregman15}.
Both the thermal Sunyaev-Zeldovich \citep[hereafter tSZ,][]{tSZ} and the kSZ effects provide alternative approaches to detect ionised gas, different to the window available from X-ray observations. The tSZ and kSZ effects provide statistical measurements of gas fraction in the visible halo population and do not restrict to the Milky Way host halo. In \cite{planck2012-XI} it is claimed that there is no apparent evidence for feedback effects in the tSZ luminosity of CGs, in contradiction with X-ray observations. Those analyses have been recently revisited by \cite{lebrunetal15} after including explicitly the impact of feedback in the filters extracting the tSZ signal. They conclude that \Planck\ tSZ observations up to $5\,R_{\rm 500}$ are  compatible with AGN-induced feedback effects, and incompatible with the no-feedback hypothesis (or self similarity in the tSZ luminosity -- halo mass relation). Those results restrict however to gas collapsed in halos that is able to generate tSZ signal \cite{Hernandezetal2006a}.  On the other hand, our kSZ study is blind to any assumed gas temperature profile in CG host halos, is not restricted to collapsed/virialised gas, and it provides {\em lower} limits to the amount of gas in CGs given the compensated structure of the AP filter. Figure~\ref{fig:fb_Dg} shows that our measurements are compatible with having detected all missing baryons in case these follow the dark matter distribution. There exist uncertainties linked to the predictions for the gas peculiar velocities, but these are relatively small if we compare the predictions from the Millennium and the Gaussian simulations in the right panel of Fig.~\ref{fig:CFs}. The impact of contaminants (such as dust or tSZ) seems to have been characterised and kept under control in \cite{Planck2015-CV}, where it is also shown that a satellite fraction (of $\sim 12\,\%$) in our CGC should decrease the kSZ amplitude by less than 10\,\%. We thus conclude that the measured kSZ signal provides evidence for all the missing baryons predicted to be inside and around the CGs, which correspond to roughly half the total amount of baryons present in the Universe at $z\simeq 0.12$ under the SDSS DR7 angular footprint. In the future, a more detailed comparison with state-of-the-art hydrodynamical numerical simulations should shed more light on these results.

\begin{acknowledgments}
C.H.-M. acknowledges the support of the {\it Ram\'on y Cajal} fellowship No. RyC-2011-08262, the {\it Marie Curie} Career Integration Grant No. 294183 and the Spanish {\it Ministerio de Econom\'\i a y Competividad} project No. AYA2012-30789. We also acknowledge useful discussions with S.D.M. White. Y.-Z.~M thanks the support of ERC starting grant No. 307209.

\end{acknowledgments}


\bibliography{kSZIII_prl,Planck_bib}

\end{document}